**Controlling of ZnO nanostructures morphology for guiding the structural and optical properties**

Amit Kumar Rana[1], Yogendra Kumar[2], Niharika Saxena[2], Rajasree Das[1,2], Somaditya Sen[1,2], Parasharam M. Shirage[1,2,*]

[1]Department of Physics, Indian Institute of Technology, Khandwa Road, Indore-452020. India

[2]Centre of Materials Science and Engineering, Indian Institute of Technology, Khandwa Road, Indore-452020. India

* Author for correspondence (E-mail: pmshirage@iiti.ac.in, paras.shirage@gmail.com)

**Abstract:**

Most of the applications of the nanostructures dependent on the morphologies which affects the opto-electronics properties. This research article provides a pathway of guiding optical properties like band-gap and fluorescence properties by controlled growth of nano-rods, -flowers, -needles or- tubes without external chemical doping, by simple hydrothermal method by controlling over synthesis parameter, temperature. Ultraviolet-visible (UV-vis) spectroscopy shows enhancement in the band gap, with increasing synthesis temperature. Photoluminescence (PL) data indicates existence of defect in the nanomaterials, which is more elaborately explained by schematic band diagram. Tuning optical properties by materials morphology is a very important parameter for device applications which is empirically demonstrated here. The sharp and strong peak in Raman Spectroscopy is observed at ~438 cm$^{-1}$, assigned to the $E_2^{high}$ optical mode of the ZnO, characteristic peak for the wurtzite hexagonal phase with highly-crystalline.

**1. Introduction:**

Controlling of the morphology of the semiconducting nanostructures is most important in the prospective of optical-electronic technological applications. ZnO has received lot of attention as a nanostructured material due to its unique properties related to numerous applications[1].It is a wide band gap semiconductor (3.3eV) with wurtzite crystal structure and high exciton binding energy of 60 meV[2,3]. The primary motivation to explore the material properties of ZnO nanostructure is its potential for a variety of practical applications, such as in opto-electronic devices (light emitting diodes (LEDs), laser diodes, solar cells, photodetectors), energy harvesting devices (Nano generators), electronic devices (transistors), sensors, catalysts, piezoelectric *etc*[4]. ZnO is also an enormous choice in research as it is often regarded as non-toxic and biocompatible, in spite of the rising evidence of significant Eco toxicity[5]. Use of the nanostructures, will allow building of nano-scale devices like nano-sensors, nano cantilevers, field-effect transistors and nano resonators for a variety of military, homeland security and commercial applications. ZnO is transparent to visible light. ZnO can easily be handled to yield uniform and high quality nanostructures and its synthesis does not require toxic precursors. All these applications of ZnO nano-structures mostly depends on the morphologies and which can be controlled be control over synthesis parameters.

A number of studies on the preparation technique of ZnO nanostructure is reported using wet chemical, electro-deposition, template assisted route, thermal evaporation, chemical vapor deposition (CVD) techniques, most of these techniques require sophisticated instruments to generate high temperature or high pressure. Hydrothermal is most simple and cost effective technique to prepare highly crystalline ZnO. To further ensure economically cheap technique, we have used amorphous glass substrates for the synthesis of nano structure. But the true challenge lies in the synthesis condition which control the crystal growth in a particular fashion to grow desired nano structures moreover prevention of cluster formation. Previously, this complex growth mechanism is mostly explain based on catalyst-driven mechanisms (solution-liquid solid[6,7] and vapor-solid-solid[8] growth) and vapor-liquid-solid (VLS)[9,10] mechanism. But recent studies shows that axial screw dislocation mechanism drive the spontaneous formation of nano wire/tube structure[11,12]. Size and morphology plays a vital role in the physical properties of the nanomaterial. So here, we have focused on the variation of optical properties with the change in shape and size of ZnO nano structures.

The objective of this report is to control morphology of ZnO nanostructure to tune structural and optical properties by controlling synthesis temperature in the hydrothermal method. Here, it is demonstrated how to grow ZnO nanostructures in the form of nano-flowers, nano-needles, nano-rods and nano-tubes just by controlling the synthesis temperature with excellent crystallinity. UV-vis, Photoluminescence and Raman measurements are performed to analyze the optical properties of the synthesized structures.

## 2. Experimental technique:

### 2.1. Synthesis Process

Zinc Acetate dehydrate (Zn $(CH_3COO)_2 \cdot 2H_2O$) (100mM: Alfa Aesar chemicals) is used as the starting precursor for preparation of pure ZnO films on glass substrate by a most simple and inexpensive hydrothermal method on a large area at atmospheric pressure. An aqueous solution of Zinc Acetate is prepared in a 100 ml double distilled water as starting material. After complete dissolution of chemicals in distilled water, ammonia is added drop wise to form resultant solution with a pH~13. The substrates are cleaned first with distilled water, followed by Acetone and Ethanol by ultra-sonication for 10 min in each solution and air dried. Glass substrate is kept vertically in the solution. To understand the growth mechanism more clearly, reaction time (120 min) was kept constant and reaction temperature is varied from 70 °C to 120 °C. Deposited films are taken out from the solution and washed with distilled water for several times to remove over deposited/ ammonia related impurities from the film. Dried films are annealed in air at 150 °C for 2 hrs to obtain pure phase ZnO.

The phase purity, morphology and composition of ZnO films are investigated by x-ray diffraction (XRD, Bruker D8 Advance X-ray diffractometer) with Cu-K$_\alpha$ radiation($\lambda$=1.54Å) and field emission scanning electron microscope (FESEM, Supra 55 Zeiss). Optical band gap is determined with the Diffuse Reflectance Spectrometer (Agilent Cary-60 UV-Vis). RT fluorescence spectroscopic measurements is conducted using spectrofluorometer (Horiba Jobin Yvon, Fluorolog-3) having Xe lamp source with an excitation wavelength of 325 nm. Micro Raman Scattering measurement of the samples is recorded using Labram-HR 800 spectrometer equipped with excitation radiation at wavelength of 488 nm from an argon ion laser at a spectral resolution of about 1 cm$^{-1}$

*Table-I Deposition temperature and sample name assignment.*

| Deposition Temperature (°C) | Sample Name (Deposition Time 120 min) |
|---|---|
| 70 | ZN1 |
| 80 | ZN2 |
| 90 | ZN3 |
| 100 | ZN4 |
| 110 | ZN5 |
| 120 | ZN6 |

**3. Results and Discussion**

Figure 1 shows XRD pattern of ZnO nano-structure deposited on glass substrate at different deposition temperatures keeping deposition time constant (120 min.). All major peaks (2Θ ~ 30° to 60°) can be assigned to the pure hexagonal phase of wurtzite-type ZnO structure with space group *P63mc*. Fig 1 shows that at such a low synthesis temperature a highly crystalline and pure phase of ZnO has been obtained. The film grown below 120°C shows a more preferred orientation along (002) planes in comparison to powder samples where (101) planes have the strongest peak intensity[13]. However the nano-material films grown at 120 °C, shows a reduced relative intensity of the (002) reflection, disappearance of (*00l*) plane is might be due to the change in structure[14]. Fig 2 shows Reitveld refinement plot of ZnO deposited at 90 °C using "Fullprof" software. Refined lattice parameter value of the nano structures are shown in Fig. 2(B).

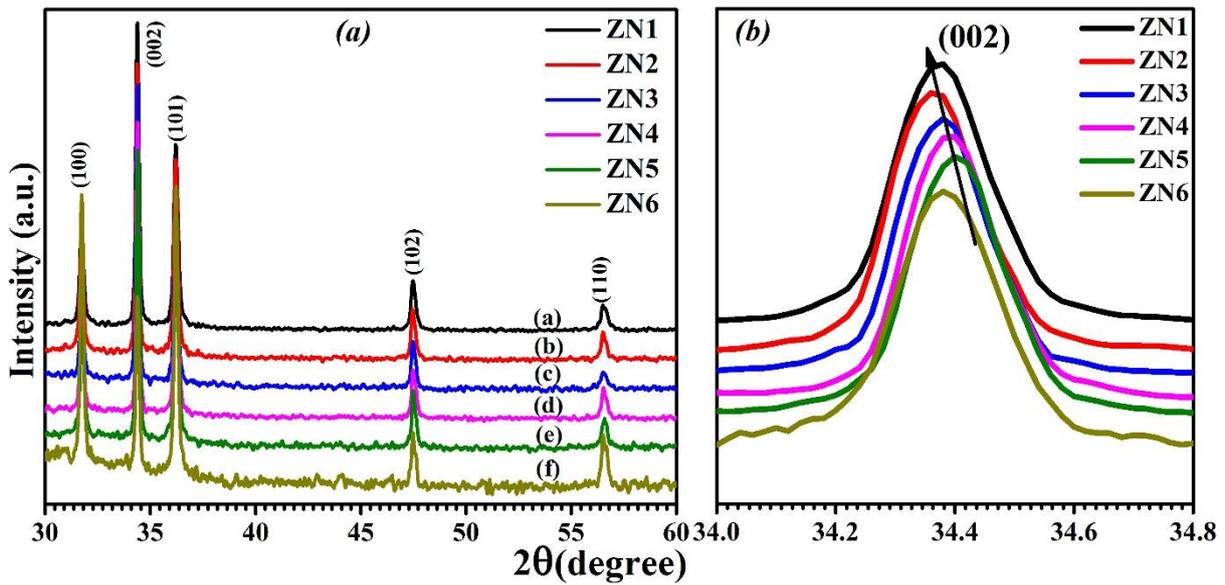

*FIG. 1 (Colour online) (a) XRD pattern of (a) ZN1 (b) ZN2 (c) ZN3 (d) ZN4 (e) ZN5 (f) ZN6 nanostructures and (b) relative shift in (002) peak position.*

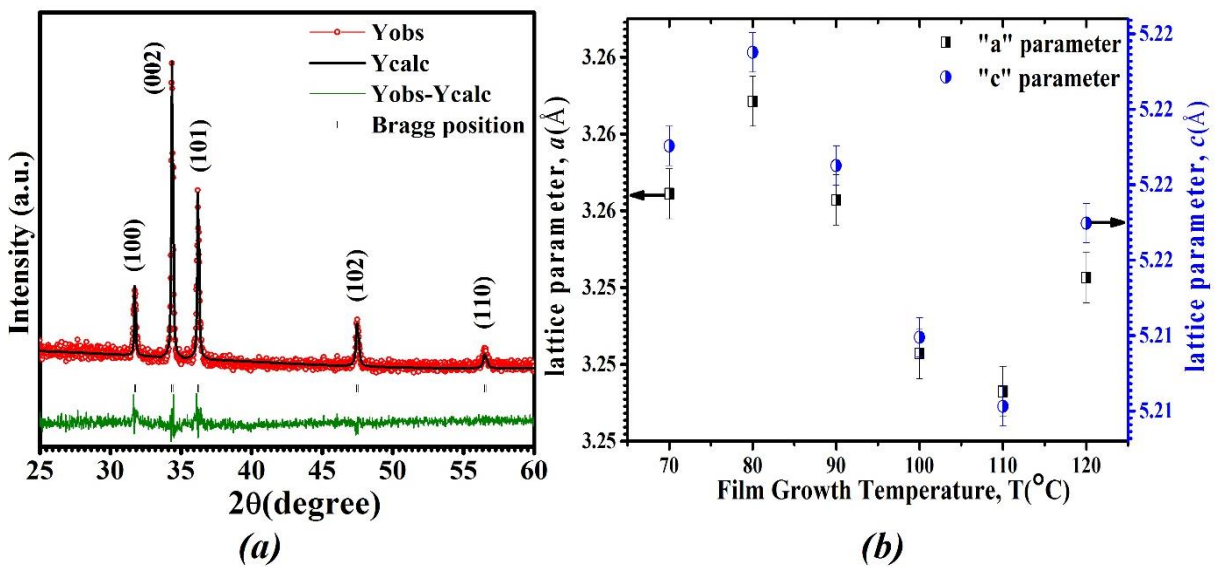

*FIG.2. (Colour online) (a) Reitveld refinement of ZnO deposited at 90°C with deposition time of 120 min. (b)The effect of deposition temperature on lattice parameter of ZnO.*

Fig 1 (b) shows (002) peaks shift towards the higher diffraction angle until growth temperature of 110 °C (ZN1 to ZN5) while for growth at 120 °C (ZN6) the peak shifts to the lower diffraction angle. This trend further verified by Reitveld refinement data, lattice constant decreases until growth temperature 110 °C followed by increase for 120 °C (Fig. 2(b)). There might be two possible reasons for the nonsystematic change in lattice parameter with growth temperature; (1) due to the change in the morphology of ZnO nano structure (2) due to internal compressive micro strain. Effect of strain developed in the nano structure grown on glass substrates were analyzed using Williamson-Hall[15] formula (Table II). Comparing Fig 2 (b) and Table II, it is clear that increase in lattice parameter has strong correlation with decrease in strain in the nanostructures.

*Table II. Shows the obtain value of crystalline size and strain (ε) for the ZnO nano structures grown at different deposition temperature.*

| Sample | Debye Scherre's method | Williamson- Hall methods Uniform Deformation Model (UDM) | |
|---|---|---|---|
| | D (nm) | D (nm) | Strain (ε) x 10$^{-3}$ |
| ZN1 | 33.21 | 84.51 | 1.53 |
| ZN2 | 37.32 | 73.72 | 0.97 |
| ZN3 | 36.72 | 107.44 | 1.72 |
| ZN4 | 36.93 | 197.18 | 2.18 |
| ZN5 | 36.03 | 206.86 | 2.33 |
| ZN6 | 35.80 | 55.66 | 1.57 |

Here we have calculated crystalline size and induced strain of the nano structures using Debye-Scherrer's formula (Eq. 1) and Williamson and Hall using uniform deformation model (UDM) (Eq. 2). Expressed as

$$D = \frac{K\lambda}{\beta_D \cos\theta} \quad \text{-----------------------------} \quad (1)$$

and

$$\beta_{hkl} = \frac{K\lambda}{D\cos\theta} + 4\varepsilon\tan\theta \qquad \text{---------------} \quad (2)$$

Where D = crystalline size, K = shape factor (0.9), λ = wavelength of $Cu_{k\alpha}$ radiation and β is full width half maxima of a corresponding peak and ε is induced strain value. All the deduced parameters are listed in Table II. From table is shows that stain is the order of $10^{-3}$ and strain strongly dependent on the crystallite size, higher the crystallite size higher is the strain and vice versa.

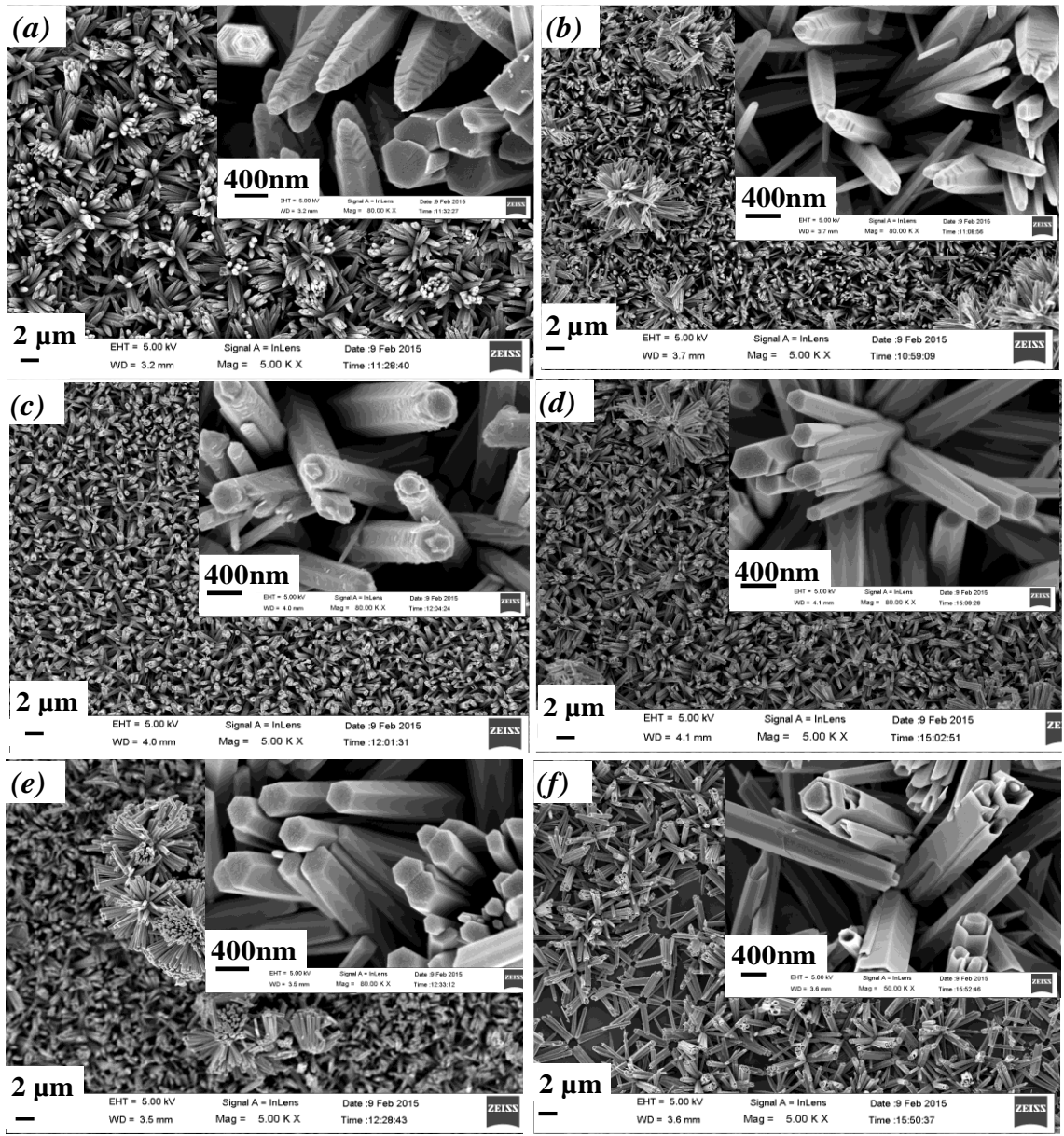

*FIG. 3. FESEM images ZnO nanostructures (a) ZN1 (b) ZN2 (c) ZN3 (d) ZN4 (e) ZN5 (f) ZN6 for deposition time of 120 min.*

Fig.3 shows FESEM images of ZnO films (magnification at 5 KX) deposited at different temperatures (70- 120 °C) and inset of the figure shows magnified image of the same sample (magnification at 80 KX).

The morphology and structure of ZnO nanostructure is not solely depend on the preparation technique, other external conditions such as ($H^+$ ion concentration) pH value, reaction temperature, reaction time, solution concentration, *etc*. also plays a vital role in hydrothermal method[16]. Comparing the nano-structure size (rods diameter) from Fig. 3, it is easily noticeable that nanorods formed at 70 °C (Fig 3 a) are bigger than the structures at 90 °C (Fig 3 b). Due to the fast nucleation process at higher temperature, bigger rods are divided/branched into smaller rods, with a decreases of size from 323 nm (ZN1) to 206 nm (ZN2) respectively. Afterwards, due to the disk shape layer formation the structure size again increases. The average size of sample from ZN1 to ZN6 is 323 nm, 210 nm, 206 nm, 228 nm, 271nm and 339 nm, respectively.

For a better understanding, schematic (Fig. 4) of idealized and proposed formation mechanism of ZnO nanostructure is drawn based on the observed nano structure in SEM. ZnO consist of an interesting structure with both polar, metastable and non-polar, stable face[17]. Zinc-terminate (001) and oxygen-terminate (00$\bar{1}$) are two polar faces (along *c*-axis) while the other non-polar faces are (010) and (110) (parallel to *c*-axis) so the growth along the (001) direction is fastest with respect to other faces (Fig. 4a) [18]. Relation velocity crystal growth along different plane is well studied by Wen-Jun Li *et al.*[19], *i.e.* $V(001) > V(01\bar{1}) > V(010) > V(011) > V(00\bar{1})$. In addition screw dislocation ( Fig. 4 *a*, 4*c* and 4*g*) and layer by layer growth (4 *b*, 4 *d*, 4 *e*) mechanisms are well-known nucleation mechanism for the growth of ZnO nano wires and tubes. In ZN2, ZN4 and ZN5, nanorods shows most possibly layer by layer growth along (00*l*) direction and form a tapered shape nano-needle on top. With the increase in reaction temperature further, the structure reduces their surface energies and tapered shape changes to flat top and gradually hexagonal rods starts appearing (Fig 3E and Fig 4(c-d)). It is also seen there are not many active sites around the ZnO nuclei at low reaction temperature. Therefore, the formed nuclei with limited growth rate can get attached together along the preferential direction to form nano-flower like structure. It is well known that crystal plane whose growth rate is slow easily appear and whose growth rate is fast easily disappear. So the (001) plane easily dissolve with respect to other non-polar face at higher temperature (120 °C) resulting the formation of nano tube (Fig. 4 *e*) [16].

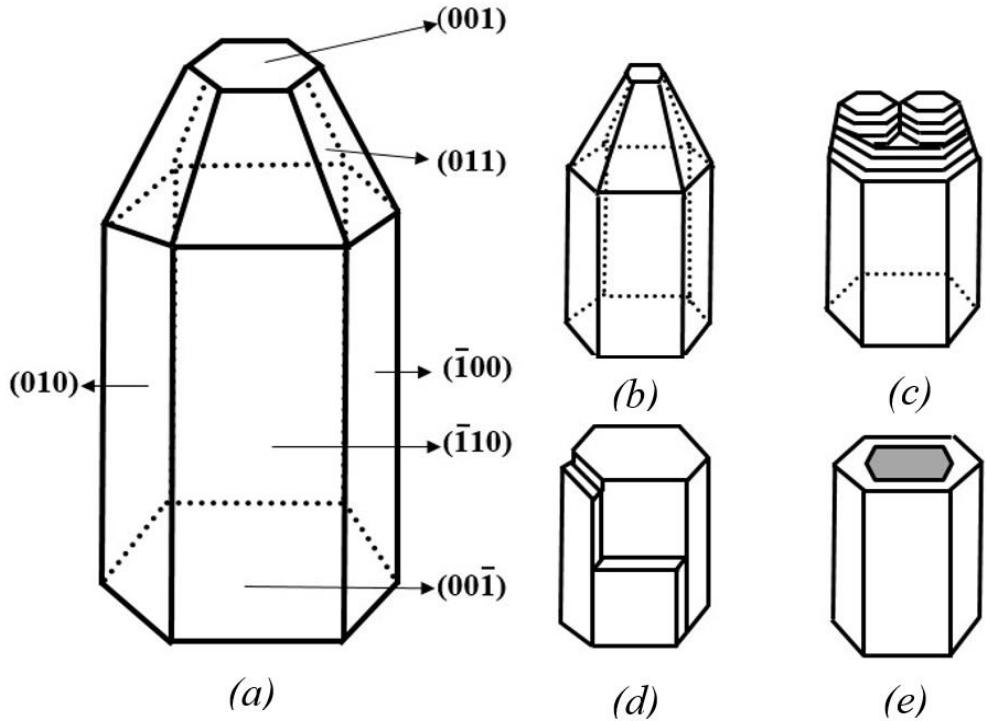

FIG.4. Schematic diagram (a) shows idealized growth process of nano-rod with tapered tip. Schematic diagram of proposed growth of (b) Nano-needle (c) Splitting of nano-rod into small nano-rods (d) layer formation along c-axis (e) Nanotube

From above Fig. 3 FESEM the growth mechanism is not clear in case of tapered nano-needle and tube formation. So we performed the experiments at fixed deposition temperatures (at 70 °C and 120 °C ) and varied the deposition time at the interval of 30 min and deposition was carried out at 30 min, 60 min, 90 min and 120 min. Fig.5 shows of FESEM images of films deposited at 70 °C for 30 min -120 min in (*a-d*) and 120 °C for 30 min -120 min from (*e-h*), it noticeably visible that the rods and tubes synthesized at different temperature and time evidenced screw dislocation mechanism as reported by Hong *et al.*, Song *et al.* and Mathews *et al.*[20-22]. Screw dislocation mechanism might responsible for modification of rods into tubes as reported by Yang *et al.*, Jin *et al.* and Bierman *et al.*[20-22]. Also one cannot rule out layer by layer deposition in these rods and tubes as from the side view of the rods and tubes one can obviously realize the layer like structure.

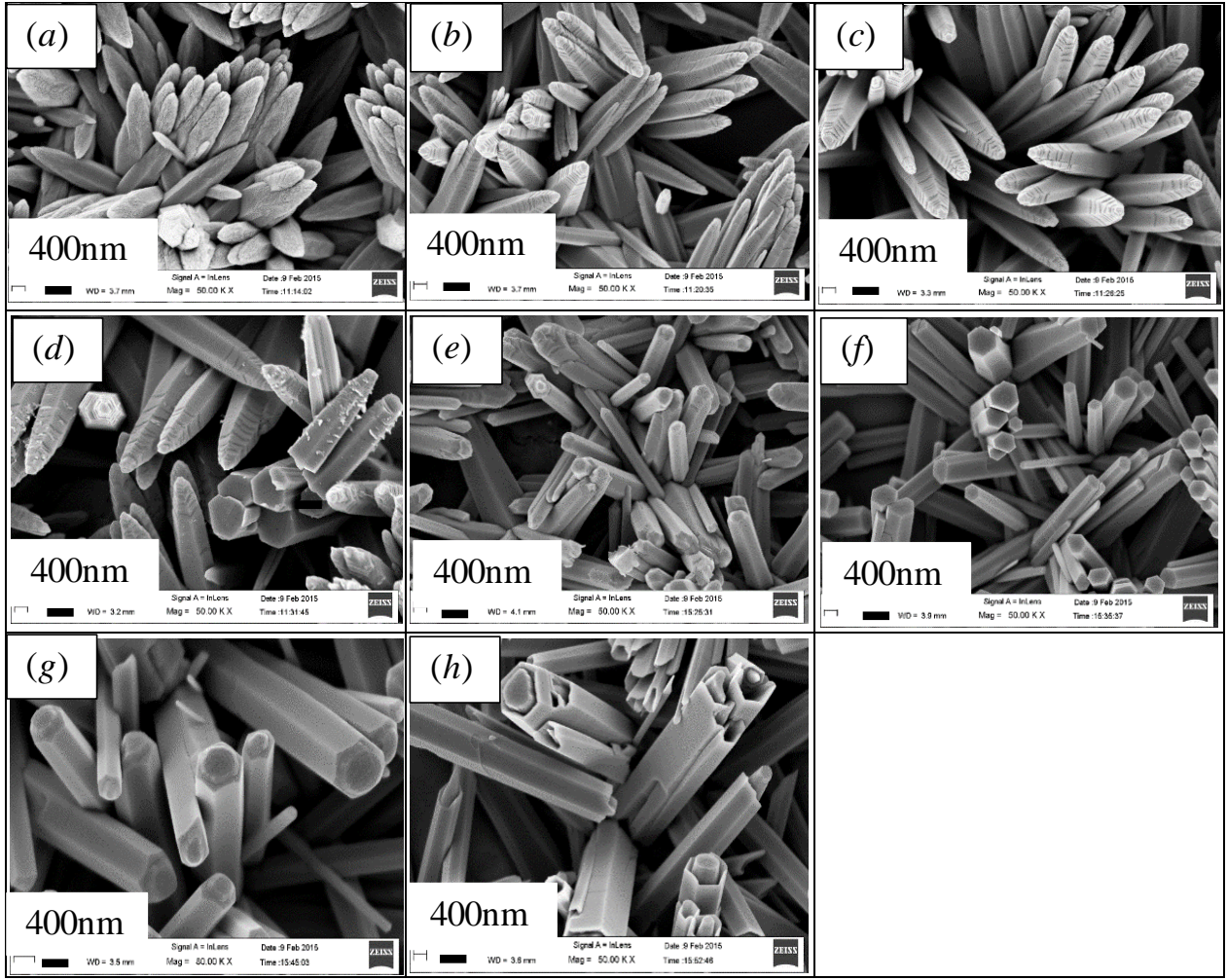

Fig 5. ZnO at 70 °C with depositing time (*a*) 30 min (*b*) 60 min (*c*) 90 min (*d*) 120 min and 120 °C with depositing time (*e*) 30 min (*f*) 60 min (*g*) 90 min (*h*) 120 min.

The effect of deposition temperature on the intrinsic optical properties of the pure ZnO were studied at room temperature (RT) by UV-Vis spectroscopic. Fig. 6. Shows the UV-visible spectra of ZnO (only shown the data in the range of (350 nm -450 nm). It is visible that the band gap is varying from 3.11eV to 3.19eV as the deposition temperature increase. The band gap was determined from Tauc plot:

$$F(R_\infty)h\nu = A(h\nu-E_g)^n$$

Where, α is the absorption constant, A is a constant, $E_g$ is the band gap value and *n* is an unit less parameter with a value 2 or ½ for indirect or direct band gap semiconductors, respectively.

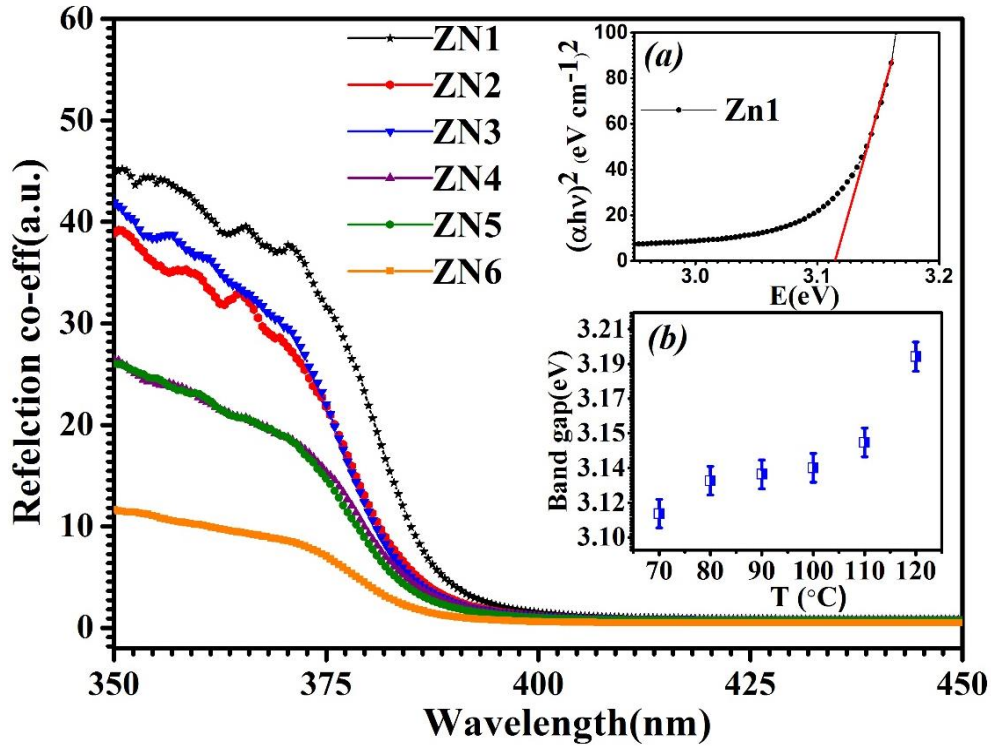

*FIG. 6. (Colour online) Room temperature UV- Vis reflection spectra of ZnO nanostructures. Inset image shows (a) Tauc plot of Zn1 and (b) Variation in bandgap with synthesis temperature.*

Inset (a) of Fig 6 shows a typical example the Tauc plot for the ZN1 sample. From the inset (b) of Fig. 6, showing the variation of band gap with growth temperature, it is very clear that the band gap is smaller (~3.11 eV) in needles ZN1, intermediate in rods (~3.14 eV) (ZN2-ZN5) and highest in tubes (~3.19 eV) ZN6. From band-gap studies it distinctly visible that the synthesis temperature play vital role which tunes the band-gap and helps to decide the device manufacturing parameter which is easily tunable in hydrothermal method by growth temperature.

Synthesis temperature dependent PL studies reveals comprehensive information about the nature of light emission and the fundamental material properties. Generally two emission bands in the PL spectrum of ZnO is observed, one is in the UV range associated with excitonic recombination or band-edge emission and another is in the visible range, which originates from the electron–hole recombination at a deep level, caused by oxygen vacancy or zinc interstitial defects[23].

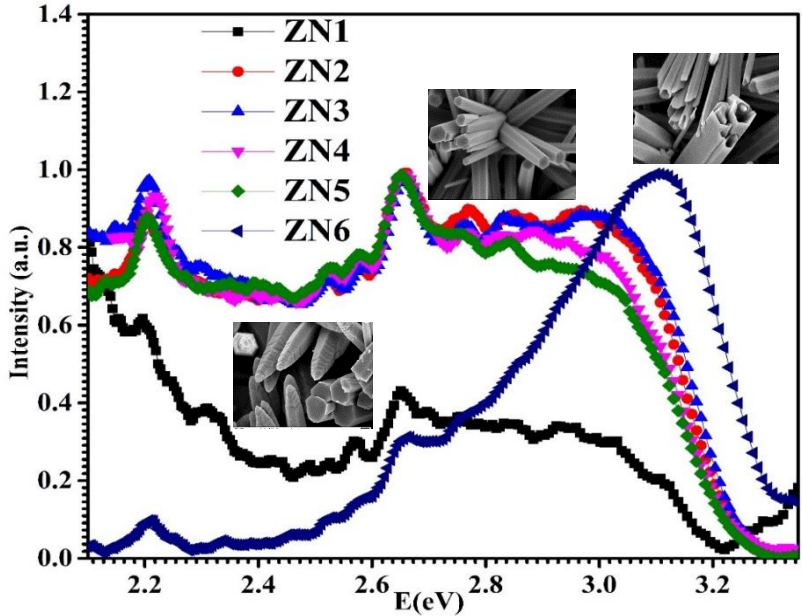

*FIG. 7 (Colour Online) Room temperature PL spectra of ZnO grown at different temperatures*

Fig. 7 shows room temperature PL emission of the samples over a broad range from 3.3 to 2.1 eV (330 to 590 nm). With increase in growth temperature of ZnO films, NBE peak position shifts toward higher energy side, irrespective of defect which is responsible for PL. The existence of Zn interstitials($Zn_{in}$), Oxygen vacancies ($V_o$), Zn vacancies ($V_{Zn}$), Oxygen interstitials ($O_{in}$), Antisite oxygen ($O_{Zn}$) have been reported as a possible defects in PL previously. Fig. 7 shows the PL spectrum of the nano-materials synthesized at different temperatures from 70-120 °C (ZN1-ZN6). The emission band is composed of a weak UV band around 3.12 nm, a weak blue band around 2.66 nm and a strong orange band around 2.20 nm. The UV emission band must be explained by a near band-edge transition of wide band gap ZnO nanorods, namely the free excitons recombination through an exciton–exciton collision process[24]. Similarly, an orange band was also observed and it was attributed to the intrinsic defect in ZnO as oxygen interstitials [25,26] suggesting oxygen excessive in the sample. We can conclude that the ZnO nanorod has a strong ability to absorb oxygen to form oxygen interstitials defects on the surface. In the case of the weak blue emission, the exact mechanism is not yet clear[27]. It may also relate to the surface defects in the present condition. From Fig. 7 it is very clear that the PL spectra strongly depends on the morphology of the ZnO nano-structures as can be seen that for needles (ZN1) intensity is low at higher energy levels. For rods it's almost flat with higher intensities, expecting certain emissions and overlap for ZN2-ZN5. For tubes (ZN6) the intensities is higher at higher energy level and then decreases. For ZnO rods (ZN2-ZN5) the yellow and green light emission is prominent from PL spectra while for ZnO nano tubes violet light. It indicates the synthesis temperature is one of the important factors to control the surface morphology which ultimately controls the optical properties. So this research paper presents the important message to the basic and applied scientists that in order to tune the optical properties the control over synthesis temperature in hydrothermal method is very important.

Fig.8 (A) shows the typical Photoluminescence spectra of ZnO nanorods grown at 90 °C measured at room temperature (excited at 325 nm) which is deconvoluted into 15 well-resolved peaks. The "■" shows experimental data. Solid green and red lines are Gaussian fitting of individual peaks and sum of all peaks, respectively. Fig .8 (B) shows the schematic band diagram. First peak $E_1$ at 3.14 eV is correspond to UV region and it is related to near band emission present in ZnO films. $E_2$ at 3.05 eV and $E_3$ at 2.92 eV is considered to appear from the swallow donor $Zn_i$ to valance band (VB) and conduction band to shallow acceptor $V_{Zn}$ respectively. Both these transitions responsible for violet-blue emissions. A broad blue peak $E_4$ at 2.82 eV in between conduction band (CB) to $O_{in}$ level and small peak $E_5$ (2.75 eV) in between $Zn_{in}$ and $V_{Zn}$, these both peak are known to originate due to the recombination of electron-hole pair.

Another intense blue peak at $E_6$ (2.65 eV) was attributed to electron transition from $Zn_{in}$ to $O_{in}$ acceptor level. The origin of green emission in ZnO is one of the most controversial issue. *Liu et al.*[28] ascribed to $Zn_i$ and $O_i$, According to Vanheusden *et al.*[29] it is related to $V_O$. Fig.8(A), shows three prominent peak $E_7$, $E_8$ and $E_9$ at 2.58, 2.53 and 2.44 eV, respectively. Calculation based on full potential linear muffin-tin orbital method explained that the position of oxygen vacancies ($V_O$) level is located at approximately 2.46 eV below the CB[30]. So these three emission are attributed to CB or deep level or trap-state oxygen singly charged $V_O$ defect state. We also observed a very strong yellow luminescence peak at 2.20 eV ($E_{13}$) and orange luminescence at 2.15 eV ($E_{14}$) below the conduction band, these defect are originated due to antisite oxygen ($O_{Zn}$) defect state and complex of $V_OZn_i$ cluster *i.e.* combination of two point defect $V_O$ and $Zn_i$, respectively[31].

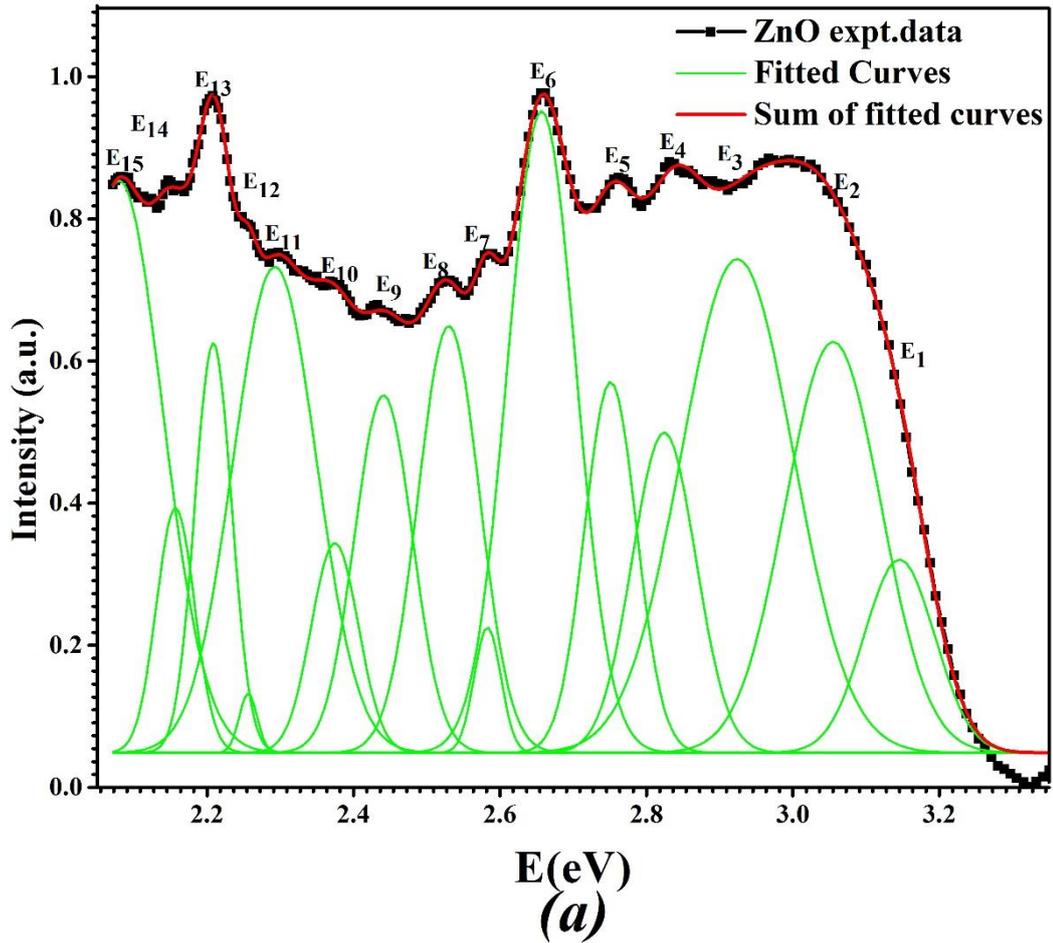

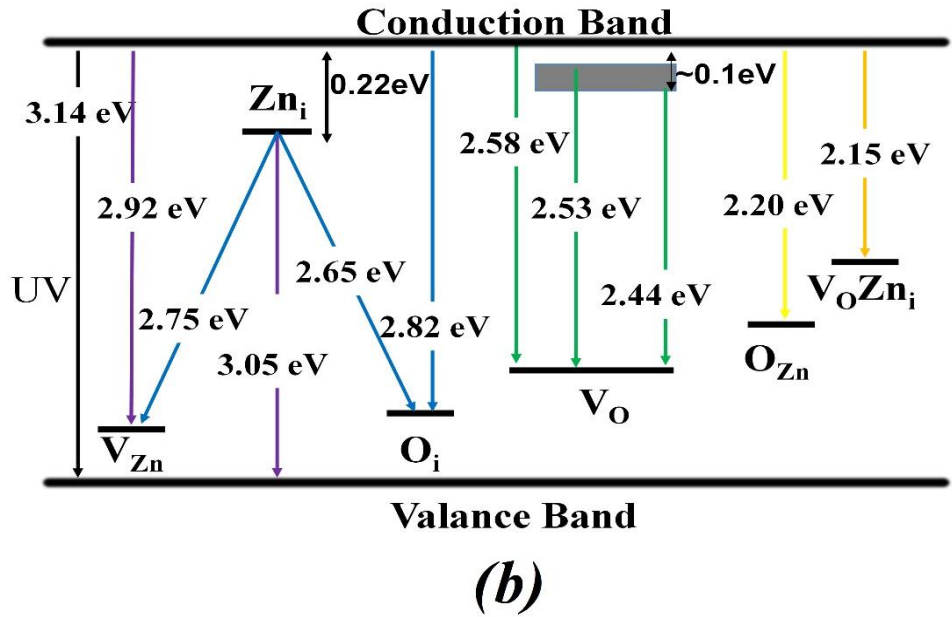

*FIG.8. (Colour online) (a) PL spectra of ZnO deposited at 90°C temperature. The "■" shows experimental data. Solid green and red lines are Gaussian fitting of individual peaks and sum of all peaks, respectively. (b) Schematic band diagram of ZnO nanostructure for ZN3.*

The optical phonon properties of the ZnO nanostructures prepared at different temperature have been investigated by Raman spectroscopy, and the results are presented in Figure 9. ZnO has a wurtzite structure that belongs to the space group $C_{6v}^4$ with two unit formulas per primitive cell, with all atoms occupying $C_{3v}$ sites[32]. The group theory predicts the existence of the following optical modes at the Γ point of the Brillouin zone: Γ = A1 + 2B1 + E1 + 2E2. ZnO has 12 branches consists of polar modes ($A_1$ and $E_1$), two non-polar modes ($2E_2$) and two silent ($2B_1$) Raman modes[33,34]. Among these, $A_1$ and $E_1$ polar modes are divided into transverse optical (TO) and longitudinal optical (LO) phonons due to the long-range electrostatic forces[32,35]. It is well understood that the electrostatic forces dominate the anisotropy in the short-range forces, the TO-LO splitting is larger than the $A_1$-$E_1$ splitting. Whereas, E2 is divided into $E_2^{low}$ and $E_2^{high}$ active modes. The $A_1$ and $E_1$ branches are both infrared and Raman active, the two non-polar $E_2$ branches are Raman active only, and the $B_1$ branches are inactive. The atoms move parallel and perpendicular to the $c$ axis, for the lattice vibrations with $A_1$ and $E_1$ symmetries, respectively. The low-frequency $E_2$ mode ($E_2^{low}$) is related with the vibration of the heavy Zn sub-lattice, while the high-frequency $E_2$ mode ($E_2^{high}$) comprises simply the oxygen atoms. According to the theory, the scarring peaks at ~ 438, 381 and 583 cm$^{-1}$, corresponding to the fundamental optical modes of E2, A1(TO), and A1(LO), respectively[14] and they have been attributed to the intrinsic defects, such as oxygen vacancy and interstitial zinc.

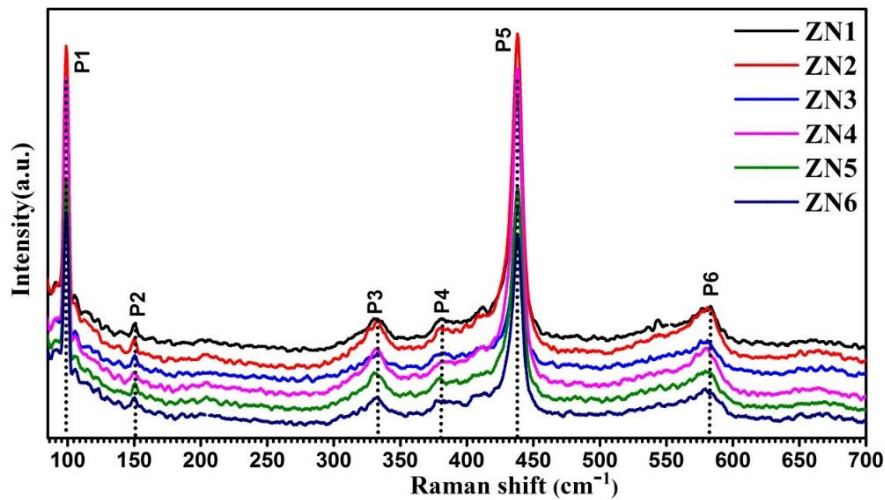

*FIG.9. RT Raman spectra of ZnO nanostructures synthesized at different temperature.*

The spectrum of all ZnO nano structure grown at different temperature is similar (Fig 9). Table III summarizes the Phonon mode frequencies (in units of cm$^{-1}$) of wurtzite ZnO nano structures.

By comparing the obtained modes of ZnO with literature, one can assign the peak around 98 cm$^{-1}$ (P1) to *$E_2$ (low)* mode. There is a secondary phonon mode has presented at 150 cm$^{-1}$, which is assigned as 2E$_2^{low}$ [36]. The small peak ~330 cm$^{-1}$ (P3) is assigned to *$E_2^{high}$ - $E_2^{low}$* (multi-phonon process) and is known to be a second order vibration mode arising from zone-boundary phonons[37]. Another small peak, P4 is observed at ~380 cm$^{-1}$ is assigned to *A1(TO)* mode. P5 around 438.1cm$^{-1}$ is assigned to $E_2^{high}$ which the characteristics of wurtzite structure and good crystallinity. Wide peak around 580 cm$^{-1}$ (P6) is assigned to the combination of A1 and E1 longitudinal optical mode, LO (A1+E1) mode. A red shift in P6 can be explained on the basis of possible mechanisms; (1) spatial confinement within the nano-rods/nano-tubes boundaries; (2) Phonon localization by the defects such as oxygen deficiency, zinc excess, surface impurities *etc*. The broadening of the spectra and strong red shifts of LO modes are reported due to the optical phonon confinement[38]. Table III shows that crystalline size of ZnO nanostructures are around 100 nm so phonon confinement might be responsible for the observed P6 peak shift.

***Table III*** *Phonon mode frequencies (in units of cm$^{-1}$) of wurtzite ZnO films deposited at different temperatures (ZN1-ZN6).*

| Peak | Position of the vibration bands (cm$^{-1}$) | | | | | | Phonon mode |
|---|---|---|---|---|---|---|---|
| | ZN1 | ZN2 | ZN3 | ZN4 | ZN5 | ZN6 | |
| P1 | 98.7 | 98.7 | 98.7 | 98.7 | 98.7 | 99.7 | *$E_2^{low}$* |
| P2 | 149.9 | 149.9 | 150.5 | 150.5 | 150.5 | 149.9 | *2$E_2^{low}$* |
| P3 | 330.1 | 333.1 | 333.1 | 332.5 | 329 | 331.8 | *$E_2^{high}$ – $E_2^{low}$* |
| P4 | 380.8 | 380.2 | 380.8 | 381.4 | 379.6 | 382.1 | *$A_1$(TO)* |
| P5 | 438.1 | 438.1 | 438.1 | 438.1 | 438.1 | 438.1 | *$E_2^{high}$* |
| P6 | 583.3 | 581.5 | 580.9 | 580.3 | 579.1 | 579.1 | *LO ($A_1 + E_1$)* |

## 4. Conclusions

Simple technique for the highly-ordered growth of ZnO nano-flowers, nano-rods and nano-tubes films on glass substrates by controlling merely growth temperature is established. The XRD studies indicates the films are highly *c*-axis oriented. Reitveld refinement confirms the first decrease in the lattice constant until growth temperature 110°C and then increase for 120°C, even though it is small. The FESEM images clearly indicates the change in the ZnO morphology with synthesis temperature, at lower temperature (70°C) needle flowers, at (80-110°C) rods-flowers and at (120°C) tube formations evidenced. The possible growth mechanism is explained. Depending on the morphologies naturally band-gap was found to contrast. PL data indicates the intensities and defect levels changes with synthesis temperatures too. The presence of defects formed in highly non-equilibrium conditions had a significant impact on the luminescence of ZnO. We demonstrated a superior control not only on the morphology but also on the defect levels for nano-rods by controlling synthesis temperature. Such ample defect level adjustments will greatly benefit the applications of ZnO nano-needles/rods/tubes in light emission, opto-electronic devices, biological labelling, display devices, *etc.*

**Acknowledgments**

This work was supported by the Department of Science and Technology, India by awarding the prestigious 'Ramanujan Fellowship' (SR/S2/RJN-121/2012) to the PMS. PMS is thankful to Prof. Pradeep Mathur, Director, IIT Indore, for encouraging the research and providing the necessary facilities. We are thankful to Dr. Vasant Sathe, IUC-DAE Consortium for Scientific Research, Indore for his help to do Raman measurement of the samples. The help received from, Dr. Anjan Chakraborty is also acknowledged. Authors are thankful to SIC Indore for providing the research facilities like XRD and FESEM.